\documentclass[prd,reprint,showpacs,showkeys]{revtex4}
\usepackage{amsfonts}
\usepackage{amssymb}
\usepackage{amsmath}
\usepackage{graphicx}
\usepackage{float}
\usepackage[font={footnotesize,it}]{caption}
\usepackage{tikz}
\usepackage{tikz-3dplot}

\begin{document}

\title{Thermodynamic Stability of a Schwarzschild Thin-Shell Wormhole}
\author{S. Danial Forghani}
\email{danial.forghani@emu.edu.tr}
\author{S. Habib Mazharimousavi}
\email{habib.mazhari@emu.edu.tr}
\author{M. Halilsoy}
\email{mustafa.halilsoy@emu.edu.tr}
\affiliation{Department of Physics, Faculty of Arts and Sciences, Eastern Mediterranean
University, Famagusta, North Cyprus via Mersin 10, Turkey}
\date{\today }

\begin{abstract}
The thermodynamic stability of a thin-shell wormhole in a Schwarzschild bulk
is considered. From the first law, entropy function is found which satisfies
the local intrinsic stability conditions. Heat capacity emerges as a
well-defined regular function justifying the stability of a Schwarzschild
thin-shell wormhole. The scope of applications of the method is not limited
by the Schwarzschild wormhole.
\end{abstract}

\maketitle

\section{Introduction}

In order to confine the indispensable exotic matter to a minimal area in a
wormhole geometry, Visser introduced the concept of thin-shell wormholes
(TSWs) \cite{Visser1} which gained much popularity in recent times \cite%
{Lobo2,Eiroa3,Yue1,Diemer1,Sharif2,Mazhari1,Sharif3}. The process involves
excising out any possible event horizon/singularity in the bulk spacetime,
applying what is called the cut-and-paste technique and obtain a new
acceptable manifold. This resulting manifold is expected to provide a safe,
traversable tunnel from one universe to the other \cite{Visser2}. Taking for
granted the negative energy density ($\sigma $) \cite{Poisson1}, we define a
surface stress-energy tensor of the form $S_{i}^{j}=diag\left( -\sigma
,p,q\right) $ for a fluid. Here $p$\ and $q$\ refer to the
tangential/angular/lateral pressures of the fluid on the shell, and in this
letter we shall have the symmetric case $q=p$.

In general relativity whenever an innovative idea is introduced,
traditionally it is applied first to the most familiar spacetime, namely the
Schwarzschild metric, and expectedly Visser followed the trend \cite{Visser3}%
. We recall that, by nature, Einstein's general relativity involves the
components of the surface stress-energy tensor to the first order in
derivative of the shell's radius with respect to the proper time of the
shell. This becomes an advantage in the sense that by defining the shell's
radius in terms of the proper time on the shell $\tau $ as $a\left( \tau
\right) $, it happens that it satisfies an equation of the form $\left(
da/d\tau \right) ^{2}+V\left( a\right) =0$ for a potential function $V\left(
a\right) $. Such an equation paves the way for the stability analysis \cite%
{Garcia1}\ provided we assume an equation of state (EoS) to incorporate the
pressure-energy relation For example, while in \cite{Lobo2}\ and \cite%
{Poisson1} a generic barotropic EoS is considered, in \cite{Bandyopadhyay1}
and \cite{Eiroa1} the EoS of a Chaplygin gas is applied. Also, Varela
discusses the assumption of a variable EoS in \cite{Varela1}. Afterwards,
with tuned parameters we plot stability islands for the oscillations of\
TSW. This scheme, however, does not apply to modified theories involving
higher order derivatives, simply because the underlying harmonic stability
equations contain derivatives higher than order two. Since this method of
dynamic stability has been used extensively in the literature \cite%
{Lobo2,Poisson1,Lemos6,Eiroa1,Eiroa2,Sharif1}\ we shall refrain from
repeating it here. Instead, in this letter we resort to the thermodynamic
stability as an alternative method \cite{Callen1}.

It has already been a long time that thermodynamic concepts have been
adapted within the geometric theory of gravitation. Black holes, Hawking
temperature, entropy, heat capacity and above all, the laws of
thermodynamics proved to have deep-rooted connections with gravity. For
instance, the phase transition of the thermodynamic system is associated
with the stability analysis of a gravitational system. Indeed, the heat
capacity, $C=T\left( \partial S/\partial T\right) $, where $T$ and $S$ refer
to the temperature and entropy, respectively, can be used as a test function
for stability. As an example, in \cite{Ma1} the concept of heat capacity is
used to discuss the stability of black holes in Einstein's and Gauss-Bonnet
gravity. Change in sign, and singularity in $C$ are both indications of an
unstable thermodynamic state. Borrowing the concepts/rules from
thermodynamics and defining entropy in a TSW, the first law (shortly $%
TdS=dM= $Mass-Energy change) becomes instrumental in the stability of a TSW.
Let us note that prior to making use of heat capacity the entropy function
itself must satisfy certain restrictions \cite{Callen1}. It is well-known
that the second law, i.e. $\Delta S\geq 0$, eliminates certain processes as
non-physical. Lemos \textit{et al.} have successfully taken advantage of this
method in various articles to investigate the stability of thin-shells in $%
2+1$-dimensional flat and BTZ spacetimes \cite{Lemos1,Lemos3},
self-gravitating electrically charged thin-shells in $3+1$-dimensions \cite%
{Lemos2}, quasiblack holes \cite{Lemos4}, and an extremal, electrically
charged thin shell in $3+1$-dimensions \cite{Lemos5}. Likewise, the entropy
function $S$ defined for a TSW must obey the local intrinsic stability
conditions. In this letter, having shown that these are all satisfied for a
Schwarzschild TSW, we employ the heat capacity test to conclude that such a
TSW is thermodynamically stable.

The organization of the letter is as follows: In section $II$\ the entropy
function $S$ is found and the thermodynamic stability of a Schwarzschild TSW
is investigated. In section $III$ we conclude the letter based on our
results.

\section{Thermodynamic Stability}

In the framework of General Relativity, calculating the static surface
energy density $\sigma $ and lateral pressure $p$ on the throat of a
spherically symmetric timelike TSW in a 4-dimensional spacetime is a routine
through the extrinsic curvature tensor formalism and have been done in
several articles \cite{Poisson1,Eiroa1,Lobo1,Forghani1}. Being expressed in
the natural units ($c=G=1$), these parameters, as functions of the mass of
the bulk spacetime $m$\ and the radius $R$\ which the TSW is constructed and
is assumed to be stable at, are given, with reference to the metric in Eq.
(3) below, by 
\begin{equation}
\sigma \left( R,m\right) =\left. \frac{-\sqrt{f(r,m)}}{2\pi r}\right\vert
_{r=R}
\end{equation}%
and%
\begin{equation}
p\left( R,m\right) =\left. \frac{r\partial f(r,m)/\partial r+2f(r,m)}{8\pi r%
\sqrt{f(r,m)}}\right\vert _{r=R}.
\end{equation}%
Here $f(r,m)$ is the metric function of the non-rotating, uncharged
spherically symmetric bulk spacetimes on the two sides of the TSW in%
\begin{equation}
ds_{\text{bulk}}^{2}=-f\left( r,m\right) dt^{2}+f^{-1}\left( r,m\right)
dr^{2}+r^{2}d\Omega ^{2}.
\end{equation}%
Note that $R$ is chosen such that it exceeds any possible horizon i.e. $%
R>r_{h}$. As can be observed from Eq.(1), the static energy density does not
satisfy the weak energy condition (WEC), acknowledging that the matter on
the TSW is indeed exotic.

To discuss the thermodynamics of such a TSW, we start by evoking the
differential form of the first law of thermodynamics 
\begin{equation}
T\left( R,M\right) dS\left( R,M\right) =dM+p\left( R,M\right) dA,
\end{equation}%
in which each thermodynamic variable is a function of the two independent
parameters, the mass $M$ and the radius $R$ of the shell. The total mass on
the wormhole's throat is defined by%
\begin{equation}
M\left( R,m\right) =A\sigma \left( R,m\right) =4\pi R^{2}\sigma \left(
R,m\right) .
\end{equation}%
We then compute%
\begin{equation}
dS\left( R,m\right) =\frac{-\beta \left( R,m\right) R}{\sqrt{f\left(
R,m\right) }}\frac{\partial f\left( R,m\right) }{\partial m}dm
\end{equation}%
for the differential of the entropy of the TSW with $\beta \left( R,m\right) 
$ being the inverse of the TSW's temperature, i.e. $\beta \left( R,m\right)
\equiv T^{-1}\left( R,m\right) $ (The Boltzmann constant is also set equal
to one i.e. $k_{B}=1$). Since the entropy is a state function it must
satisfy the integrability condition 
\begin{equation}
\frac{\partial }{\partial R}\left( \frac{-\beta R}{\sqrt{f}}\frac{\partial f%
}{\partial m}\right) _{m}=0,
\end{equation}%
which leads to the direct solution%
\begin{equation}
\beta \left( R,m\right) =-\frac{\sqrt{f\left( R,m\right) }}{R\left( \partial
f/\partial m\right) }B\left( m\right)
\end{equation}%
with $B\left( m\right) $ being any arbitrary function of mass $m$. This
solution, in turn, directs us to 
\begin{equation}
S\left( m\right) =\int B\left( m\right) dm+S_{0}
\end{equation}%
where $S_{0}$ is an integration constant. The simplest form for the function 
$B\left( m\right) $ would be a power function of the form 
\begin{equation}
B\left( m\right) =\alpha m^{\gamma },
\end{equation}%
which leads us to the expression for the entropy as follows%
\begin{equation}
S\left( m\right) =\left\{ 
\begin{array}{c}
\frac{\alpha }{\gamma +1}m^{\gamma +1}+S_{0}\text{ \ \ \ \ }\gamma \neq -1
\\ 
\alpha \ln \left( m\right) +S_{0}\text{\ \ \ \ }\gamma =-1%
\end{array}%
\right. .
\end{equation}%
(Note that he logarithmic solution is normally expressed as $\alpha \ln
\left( m/m_{0}\right) $, with $m_{0}$\ being a constant to fix the
dimension. However, we prefer to keep it as it is in Eq. (11) with $%
S_{0}=-\alpha \ln \left( m_{0}\right) $,\ so that we can treat the solutions
on an equal footing). Let us not forget that through Eq.(5), the entropy is
in fact in terms of the independent variables $R$\ and $M$. Before we
proceed, let us have a quick discussion on the coefficient $\alpha $. The
requirement of entering $\alpha $ into the calculations via Eq.(10) is to
keep the temperature of the TSW and the entropy physically meaningful. More
specifically, by setting the sign of $\alpha $ we demand that $\beta $ and
hence the temperature takes on positive values only. Originally, $\alpha $
is merely a scale factor and except for this positivity obligation it is of
no importance and therefore without loss of generality one may set it equal
to unity or any other value in order to simplify the equations to the best.
(Nonetheless, $\alpha $\ is not dimensionless and its proper unit must be
kept.)

The next step would be to examine whether the stability conditions \cite%
{Callen1,Lemos3}%
\begin{equation}
\left\{ 
\begin{array}{c}
\left( \frac{\partial ^{2}S}{\partial M^{2}}\right) _{R}\leq 0 \\ 
\left( \frac{\partial }{\partial R}\left( R^{-1}\frac{\partial S}{\partial R}%
\right) _{M}\right) _{M}\leq 0 \\ 
\left( \frac{\partial ^{2}S}{\partial M^{2}}\right) _{R}\left( \frac{%
\partial }{\partial R}\left( R^{-1}\frac{\partial S}{\partial R}\right)
_{M}\right) _{M}-R^{-1}\left( \frac{\partial ^{2}S}{\partial M\partial R}%
\right) \geq 0%
\end{array}%
\right. ,
\end{equation}%
hold for this entropy or not. To do this let us be more specific by
considering a TSW constructed from two Schwarzschild spacetimes, for which%
\begin{equation}
f\left( R,m\right) =1-\frac{2m}{R},
\end{equation}%
and therefore%
\begin{equation}
\beta \left( R,R_{s}\right) =R_{s}^{\gamma }\sqrt{1-\frac{R_{s}}{R}},
\end{equation}%
and%
\begin{equation}
S\left( R_{s}\right) =\left\{ 
\begin{array}{c}
\frac{R_{s}^{\gamma +1}}{\gamma +1}+S_{0}\text{ \ \ \ \ }\gamma \neq -1 \\ 
\ln \left( R_{s}\right) +S_{0}\text{ \ \ \ \ }\gamma =-1%
\end{array}%
\right. ,
\end{equation}%
where $R_{s}=2m$ is the Schwarzschild radius of the original spacetimes.
Note that now $\alpha >0$, and it is readjusted so that we can obtain the
simple forms in the two latter equations. It is also worth noting that the
expression for $\beta $ in Eq. (14) is in full agreement with Tolman formula
with $T_{\infty }=R_{s}^{-\gamma }$ \cite{Tolman1}.\ Furthermore, the TSW is
expected to become hotter by setting it closer to the event horizon,
therefore with an increase in the mass of the black hole (and so in $R_{s}$)
we expect an increase in the temperature. This implies that we must
disregard the positive values of $\gamma $ and take $\gamma \leq 0$ as the
admissible domain for the power. Since at the limit $R_{s}\rightarrow 0$ we
expect $S=0$, we can choose $S_{0}=0$ and also $\gamma >-1$ (The logarithmic
solution is not valid). Finally, the entropy can be expressed in terms of $M$
and $R$ by replacing $R_{s}$ with $2m$, where $m$ itself is related to $M$
and $R$ through Eq. (5). By changing $\gamma +1\rightarrow \zeta $, we
finally arrive at%
\begin{equation}
S\left( R,M\right) =\zeta ^{-1}\left( R-\frac{M^{2}}{4R}\right) ^{\zeta };%
\text{ \ \ \ \ }0<\zeta \leq 1,
\end{equation}%
which has an appropriate form to be investigated under the stability
conditions. The three conditions in Eq. (12) then read%
\begin{equation}
\left\{ 
\begin{array}{c}
-\frac{R_{s}^{\zeta -2}}{2R}\left[ R_{s}+2\left( 1-\zeta \right) \left(
R-R_{s}\right) \right] \leq 0 \\ 
-\frac{R_{s}^{\zeta -2}}{R^{3}}\left[ \left( 1-\zeta \right) \left(
2R-R_{s}\right) +R_{s}\left( 4R-3R_{s}\right) \right] \leq 0 \\ 
\frac{R_{s}^{2\zeta -3}}{2R^{4}}\left[ 4\left( 1-\zeta \right) \left(
R-R_{s}\right) ^{2}+\left( 2-\zeta \right) \left( 2R-R_{s}\right) R_{s}%
\right] \geq 0%
\end{array}%
\right. ,
\end{equation}%
which are all immediately satisfied (recalling that \ $0<\zeta \leq 1$\ and $%
R>R_{s}$), with no further restrictions, implying that the TSW is
thermodynamically stable. Let us note that the scaling property of this
entropy function is marked by $S\rightarrow \lambda ^{\zeta }S$, as $%
R\rightarrow \lambda R$ and $M\rightarrow \lambda M$ for a constant $\lambda 
$.

Furthermore, we can assess the heat capacity of the matter on the TSW to see
if it is reasonably physical or not. To do so, we summon%
\begin{equation}
C=T\left( \frac{\partial S}{\partial T}\right) _{R},
\end{equation}%
where $C$ is the heat capacity of the matter on the TSW. By using the chain
rule and Eqs. (5), (14) and (16) one easily obtains%
\begin{equation}
C=\frac{2R_{s}^{\zeta }\left( R-R_{s}\right) }{2\left( 1-\zeta \right)
\left( R-R_{s}\right) +R_{s}},
\end{equation}%
which is evidently positive definite and physical.

\section{Conclusion}

Since Visser's articles in 1989 \cite{Visser1,Visser3}, in which a practical
approach to investigate the stability analysis of a radially perturbed TSW
was proposed, the stability of TSWs has been studied vastly for almost four
decades. Not to mention that this approach is still the authors' favorite
and is being applied more than any other method \cite%
{Lobo2,Eiroa2,Sharif1,Dias1}. While in this procedure the determining factor
is the dynamics of the TSW, here we followed a rather different approach in
which the thermodynamic stability of the TSW is investigated in its static
status; if the TSW satisfies the first law of thermodynamics such that it
passes the thermodynamic stability conditions successfully, and gives rise
to a meaningful heat capacity, then it can be considered stable, in
thermodynamic sense. This method has already been applied successfully to
thin-shells \cite{Lemos1,Lemos3,Lemos2,Lemos4,Lemos5}, but to the best of
our knowledge not to TSWs. Our aim with this letter was to fill this gap. To
sum up, it can be stated that a static TSW, constructed out of a
Schwarzschild bulk spacetime, can be thermodynamically stable. Let us add
that thermodynamic stability can be considered as a new test method for TSWs
in theories that the standard dynamic method fails to work. For instance,
Gauss-Bonnet (or more generally the Loveloce theory) and massive gravity
theories whose actions contain higher order derivatives are in this
category. In analogy with the case of wormholes \cite{Pedro1},
thermodynamics of TSWs can be extended to cover non-equilibrium phenomena,
such as disintegration and formation of baby TSWs. We add finally that
application of both the dynamical (in case that it works) and
thermodynamical methods provides a complementary analysis for the stability
of a TSW.

\end{document}